\documentclass[conference]{IEEEtran}
\IEEEoverridecommandlockouts
\usepackage{lipsum}
\usepackage{cite}
\usepackage{flushend}
\usepackage{amsmath,amssymb,amsfonts}
\usepackage{graphicx}
\usepackage{textcomp}
\usepackage{xcolor}
\usepackage[utf8]{inputenc}
\usepackage[T1]{fontenc}
\usepackage{hyperref} 
\usepackage{algorithm}
\usepackage{algorithmic}
\usepackage{comment}
\usepackage{threeparttable}
\usepackage{subcaption}
\usepackage{array}
\usepackage{caption}
\usepackage{booktabs}
\usepackage{multirow}
\hypersetup{colorlinks=true, linkcolor=blue, filecolor=magenta, urlcolor=cyan, citecolor=blue}

\usepackage{url}

\def\BibTeX{{\rm B\kern-.05em{\sc i\kern-.025em b}\kern-.08em
    T\kern-.1667em\lower.7ex\hbox{E}\kern-.125emX}}

\makeatletter
\newcommand\fs@betterruled{%
  \def\@fs@cfont{\bfseries}\let\@fs@capt\floatl@ruled
  \def\@fs@pre{\vspace*{6pt}\hrule height.8pt depth0pt \kern2pt}%
  \def\@fs@post{\kern2pt\hrule\relax}%
  \def\@fs@mid{\kern2pt\hrule\kern2pt}%
  \let\@fs@iftopcapt\iftrue}
\floatstyle{betterruled}
\restylefloat{algorithm}
\makeatother

\renewcommand{\baselinestretch}{1}

\begin{document}

\title{Handover Analysis for Vehicular Communication with Explainability on the Fly}

 \author{
    \IEEEauthorblockN{Ali Fuat Sahin\IEEEauthorrefmark{1}, Semiha Tedik Başaran\IEEEauthorrefmark{1}, Tufan Kumbasar\IEEEauthorrefmark{2} }
    \IEEEauthorblockA{\IEEEauthorrefmark{1}Faculty of Electrical and Electronics Engineering, Istanbul Technical University, Istanbul, Turkiye}
    \IEEEauthorblockA{\IEEEauthorrefmark{2}AI and Intelligent Systems Laboratory, Istanbul Technical University, Istanbul, Turkiye}
    \IEEEauthorblockA{Email: \{sahinal18, tedik, kumbasart\}@itu.edu.tr}
}  

\maketitle

\begin{abstract}
Handover (HO) management in vehicular networks requires fast and reliable decision-making under highly dynamic conditions. While machine learning (ML) approaches can improve HO detection by capturing complex relationships among various key performance indicators (KPIs), their black-box nature limits interpretability and operator trust. To address this, this paper investigates HO detection from an explainability-on-the-fly perspective using inherently interpretable models based on the functional analysis of variance (fANOVA) framework. The proposed models are evaluated using two real-world operator datasets and compared against a Long Short-Term Memory baseline augmented with post-hoc SHAP explanations. Unlike post-hoc approaches, the proposed framework enables immediate interpretation of model decisions without incurring additional computational overhead. This capability is particularly critical for latency-sensitive vehicular networks. The results show that fANOVA-based models achieve competitive detection performance while providing significantly reduced explanation latency compared to conventional post-hoc methods. Furthermore, feature ranking and visualization analyses reveal physically meaningful relationships between KPIs and HO occurrences that align with standardized HO mechanisms. These results demonstrate that inherently interpretable models provide an efficient and transparent solution for HO detection in next-generation vehicular networks.
\end{abstract}


\section{Introduction}
The evolution of next-generation (NextG) wireless networks is driven by the need to support ultra-reliable, low-latency communication and enhanced data rates for emerging applications such as connected and autonomous vehicles, immersive extended reality (XR), and mission-critical services \cite{6G}. These applications impose stringent requirements on network design, particularly in highly dynamic environments. At the same time, increasing network densification, heterogeneous deployments, and dynamic user behavior introduce significant challenges in maintaining consistent quality of service. To address these demands, AI-native network design has emerged as a promising paradigm for enabling intelligent and adaptive network control \cite{aiNative}. However, these advancements also introduce complex and highly dynamic operating conditions, particularly under high mobility scenarios. In such environments, rapid fluctuations in radio conditions and frequent transitions between cells significantly complicate mobility management, requiring fast and reliable decision-making under uncertainty.

To address the mobility challenges, handover (HO) mechanisms serve as a fundamental component of mobility control in cellular networks, ensuring seamless connectivity across coverage areas \cite{handoverMetricSelection}. As network architectures evolve, HO techniques have been continuously refined to meet stringent performance requirements. Hence, a broad range of HO optimization approaches, including stochastic modeling, game-theoretic frameworks, fuzzy logic–based, and learning-based techniques have been investigated \cite{handoverSurvey, surveyMachineLearning}. In particular, machine learning (ML)-based approaches can capture complex, non-linear relationships among network key performance indicators (KPIs), enabling adaptive HO decisions. However, the decision-making process remains largely opaque, offering limited insight into which KPIs drive handover events. This lack of transparency poses a critical challenge for network operators in interpreting and optimizing the mobility behavior. Lastly, beyond prediction performance and interpretability, the heterogeneous and time-varying nature of vehicular networks introduces learning latency as another critical consideration, since frequent model updates and rapid training cycles become necessary to adapt to evolving network conditions.

To address the lack of transparency in conventional ML models, Explainable Artificial Intelligence (XAI) has emerged as a promising approach for improving interpretability in data-driven systems \cite{xai}. XAI provides human-understandable insights by quantifying the feature contributions and revealing the input-output relationships, hence enhancing trust and accountability \cite{xai6G}. In the context of wireless communications, XAI has been applied to various tasks such as KPI analysis \cite{afs2} and anomaly detection \cite{11555086}. However, most existing XAI approaches are applied in a post-hoc manner, generating explanations after model decisions are made. This limits their applicability in real-time operations, particularly in highly dynamic vehicular networks, where rapid and frequent HO decisions require immediate and interpretable insights \cite{latestXAIforHO}. 

Motivated by these challenges, this paper investigates HO detection in vehicular 5G NR networks from an explainability-on-the-fly perspective. We employ inherently interpretable models based on the functional analysis of variance (fANOVA) framework and compare them with a conventional ML model enhanced by post-hoc explanations. The study is conducted using real-world operator measurement data to capture realistic HO behavior. Our analysis provides a comprehensive evaluation of the trade-offs between detection performance, interpretability, and explanation latency. The results demonstrate that intrinsically interpretable models achieve competitive predictive performance while significantly reducing explanation latency compared to post-hoc methods. Moreover, the proposed framework enables direct identification of the most influential KPIs and reveals their underlying relationships with HO occurrence, offering physically meaningful insights into mobility behavior. These findings highlight the potential of on-the-fly explainability to enhance transparency, support operator decision-making, and improve the reliability of mobility management in next-generation vehicular networks.

\begin{figure}
    \centering
    \includegraphics[width=1\linewidth]{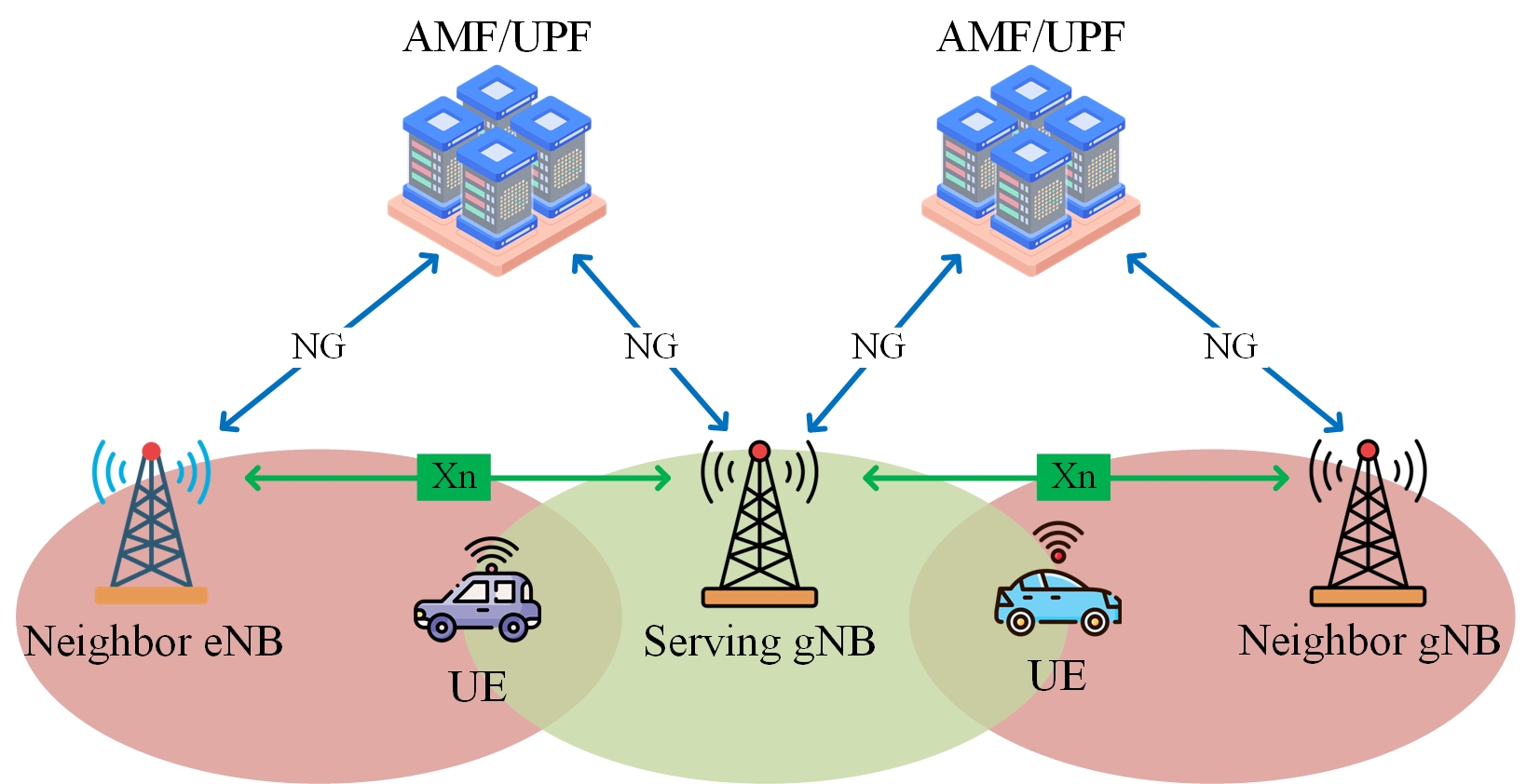}
    \caption{Handover in 5G NR.}
    \label{fig:SystemModel}
    \vspace{-15 pt}
\end{figure}

\section{System Model}
\label{systemModel}
\vspace{-2 pt}

The considered system model is illustrated in Fig. \ref{fig:SystemModel}. A 5G NR base station (gNB) serves multiple user equipments (UEs) in the network, while a neighboring gNB and a neighboring LTE base station (eNB) are also present. The 5G Core Network (CN) is responsible for overall network management and control. In particular, the Access and Mobility Management Function (AMF) handles mobility-related procedures such as HO management and paging, whereas the User Plane Function (UPF) manages user connectivity. Communication between the CN functions (AMF/UPF) and the base stations (gNB/eNB) is realized through the Next Generation Application Protocol (NGAP). Specifically, NG-C and NG-U provide the control and user plane functionalities between the base stations and AMF/UPF, respectively. Lastly, the Xn interface enables the exchange of control messages between base stations. The considered scenario focuses on vehicular communications, where UEs experience high mobility and rapidly varying channel characteristics. In such environments, HO procedures become particularly critical, as delayed HO decisions may lead to service interruption and performance degradation.

According to 3GPP, HO decisions in 5G NR are governed by predefined measurement events \cite{handoverStandard}. Radio Access Technology (RAT) refers to the underlying cellular technology, such as 5G NR or 4G LTE. The A-series events support mobility within the same RAT, whereas the B-series events are associated with mobility between different RATs. Among these, events A3, A5, and B2 are commonly considered in practical mobility management \cite{handoverEvent}. Event A3 is typically used for intra-frequency handovers, event A5 for inter-frequency handovers, and event B2 for inter-RAT handovers from a gNB to an eNB. Their trigger conditions are given as
\begin{subequations}
    \label{eq:HO_events}
    \begin{align}
        \text{A3 HO:} \quad &M_{NC} - Hys > M_{SC} + Off,
        \label{eq:A3}\\
        \text{A5 HO:} \quad &M_{SC} + Hys < T_1 \land M_{NC} - Hys > T_2,
        \label{eq:A5}\\
        \text{B2 HO:} \quad &M_{SC} + Hys < T_1 \land M_{NiRC} - Hys > T_2,
        \label{eq:B2}
    \end{align}
\end{subequations}
where $M_{SC}$, $M_{NC}$, and $M_{NiRC}$ denote the measured performance metrics of the serving NR cell (SC), neighboring NR cell (NC), and neighboring inter-RAT cell, respectively. Representative metrics include Reference Signal Received Power (RSRP), Received Signal Strength Indicator (RSSI), and Reference Signal Received Quality (RSRQ) \cite{afs2}, which are expressed in either decibels (dB) or decibel-milliwatts (dBm). The predefined thresholds $T_1$ and $T_2$ have the same units as their corresponding metrics, whereas the hysteresis $Hys$ and offset $Off$ parameters are expressed in dB.

\section{Explainable Models \& Handover Datasets}
\label{models}
This section presents the explainable models and HO datasets considered in the HO analysis. Inherently interpretable models are employed to investigate the underlying behavior of the HO process, while a Long Short-Term Memory (LSTM) classifier is trained as a predictive baseline. Considering the temporal characteristics of HO events,the LSTM classifier for $P$ KPIs and a sequence length of $T$ is formulated as
\begin{subequations}
    \label{eq:lstmClassifier}
    \begin{align}
        g(\mathbf{X}) &= \sigma\!\left(\mathbf{w}_{o}^{\mathsf{T}}\mathbf{h}_{T}(\mathbf{X})+b_{o}\right), \quad \mathbf{X}=[\mathbf{x}_{1},\ldots,\mathbf{x}_{T}],
        \label{eq:lstmOutput}\\
        (\mathbf{h}_{t},\mathbf{c}_{t}) &= f_{\mathrm{LSTM}}\!\left(\mathbf{x}_{t}, \mathbf{h}_{t-1}, \mathbf{c}_{t-1} \right), \quad t=1,\ldots,T,
        \label{eq:lstmState}
    \end{align}
\end{subequations}
where $\mathbf{X} \in \mathbb{R}^{P\times T}$ denotes the input sequence matrix, $\mathbf{x}_t \in \mathbb{R}^{P}$ is the feature vector, and $\mathbf{h}_t$ and $\mathbf{c}_t$ represent the corresponding hidden and cell states, respectively. Moreover, $\mathbf{w}_o$ and $b_o$ denote the output-layer weight vector and bias, while $\sigma(\cdot)$ is the sigmoid activation function that maps the final hidden state to the predicted HO probability, $g(\mathbf{X})$. SHapley Additive exPlanations (SHAP) is adopted as the representative post-hoc method due to its established use in wireless communications, model-agnostic nature and feature-ranking capability \cite{11555086,shap}.

\subsection{Explainable Models}
All built-in explainable models considered in this study fall under the fANOVA framework, which decomposes the response into additive terms of increasing order as
\begin{equation}
    g(x) = \sum_{j} g_{j}(x_{j})
    + \sum_{j \neq k} g_{jk}(x_{j}, x_{k})
    + \cdots,
    \label{eq:fanovaModel}
\end{equation}
where $g_j(\cdot)$ denotes the first-order (main effect) contribution of feature $x_j$ and $g_{jk}(\cdot,\cdot)$ represents the second-order interaction between features $(x_j)$ and $x_k$. Higher-order interaction terms $(g_{jkl}(x_j,x_k,x_l),\ldots)$ can be included with the trade-off between model complexity and interpretability. Accordingly, we implement three representative approaches, namely Generalized Additive Models (GAMs) \cite{gam}, Explainable Boosting Machines (EBMs) \cite{ebm}, and GAMI-Net \cite{gaminet}. A key advantage of these models is their inherent interpretability, achieved through explicit modeling of main effects and interactions, as well as through feature ranking and selection.

\subsubsection{GAM} GAMs are one of the simplest fANOVA models, utilizing the flexible smooth terms to capture nonlinearities in the data while retaining interpretability. The mathematical expression of GAMs can be given as
\begin{equation}
    g\bigl(\mathbb{E}[y]\bigr) = \mu + \sum_{i} f_{i}(x_{i}),
    \label{eq:gamModel}
\end{equation}
where $g(\cdot)$ is the link, $\mu$ the intercept, and $f_i(\cdot)$ the smooth function for the $i$th main effect. The main disadvantage of GAMs is that they can only model main effects and cannot capture higher-order terms \cite{gaam}. As data complexity grows, this limitation can impact performance, motivating more advanced models.

\subsubsection{Explainable Boosting Machines}
GAMs with Interactions (GA$^2$Ms) \cite{gaam} augment GAMs with pairwise interaction terms, using smooth functions to represent both nonlinear main effects and salient interactions with the given expression of
\begin{equation}
    g\bigl(\mathbb{E}[y]\bigr)
    = \mu + \sum_{i} f_{i}\bigl(x_{i}\bigr) 
    + \sum_{i, j} f_{ij}\bigl(x_{i}, x_{j}\bigr).
    \label{eq:gaamModel}
\end{equation}
Here, $g(\cdot)$ is the link, whereas $\mu$ is the intercept term. $f_i(\cdot)$ is the $i$th main effect, and $f_{ij}(\cdot)$ is the interaction pair for $i \neq j$. Building on the GA$^2$M framework, EBM has been proposed \cite{ebm}. EBMs are specialized gradient-boosting models that learn spline-like shape functions with smoothness constraints and regularization. EBMs can achieve competitive accuracy while providing one-dimensional plots for main effects and two-dimensional heatmaps for pairwise interactions.

\begin{table}[t]
\centering
\caption{Dataset Properties}
\vspace{-5 pt}
\label{tab:datasetProperties}
\resizebox{0.5\textwidth}{!}{%
\begin{tabular}{|c|l|l|}
\hline
\multicolumn{1}{|c|}{\textbf{Features}} & \multicolumn{2}{c|}{\textbf{Dataset Name}} \\
\hline
& Beyond Throughput & DoNext  \\
\hline
Number of Samples & 2508 & 152,775 \\
\hline
Number of Features & 25 & 28 \\
\hline
Number of Handovers (\%) & 873 (34.81\%) & 59,472 (38.93\%) \\
\hline
Cellular Technology & 5G NR $\&$ LTE & 5G NR \\
\hline
SC/NC Metrics & Yes & No (Only SC) \\
\hline
\end{tabular}
}
\vspace{-15pt}
\end{table}

\subsubsection{GAMI-Net}
While EBMs bring GA$^2$M structure into a boosted, glass-box model, they do not take advantage of advanced deep-learning methods. GAMI-Net \cite{gaminet} addresses this issue by encoding the fANOVA expression directly into the neural architecture. Explicitly, dedicated sub-networks model each main effect and each selected interaction, and their outputs are combined additively. In addition, GAMI-Net incorporates three interpretability constraints to enhance predictive performance while maintaining transparency: sparsity prunes nonessential features, heredity permits an interaction only when at least one of its parent main effects is included, and marginal clarity promotes orthogonality between main effects and their corresponding interactions. Following these constraints, the formula for GAMI-Net can be given as
\begin{equation}
    g\bigl(\mathbb{E}[y]\bigr)
    = \mu
    + \sum_{i \in S_{1}} f_{i}\bigl(x_{i}\bigr)
    + \sum_{(i,j) \in S_{2}} f_{ij}\bigl(x_{i}, x_{j}\bigr),
    \label{eq:gaminetModel}
\end{equation}
where only active main effects \(S_1\) and interaction pairs \(S_2\) are retained. The notation \(\mu\), \(f_i(\cdot)\), and \(f_{ij}(\cdot)\) is consistent with \eqref{eq:gaamModel}. For further details, see \cite{gaminet}. 

\subsection{Handover Datasets}
As shown in (\ref{eq:A3},~\ref{eq:A5},~\ref{eq:B2}), HO analysis requires radio-link metrics for both the SC and NC simultaneously. However, many datasets in the literature emphasize network-level metrics rather than radio-link performance, and most HO studies rely on simulation data instead of real-world measurements. To address these limitations, two publicly available datasets are employed \cite{commDataset, doNextDataset}. Using these datasets, a windowing approach is applied to obtain samples suitable for HO analysis. For both datasets, the target label is derived from changes in the SC index: it is set to 1 when consecutive samples indicate different SCs, thereby representing the HO events.

\begin{table}[t]
\centering
\caption{The KPIs for the Handover Analysis}
\vspace{-5 pt}
\label{tab:kpi}
\begin{tabular}{|p{3.2 cm}|p{4.8 cm}|}
\hline
\multicolumn{1}{|c|}{\textbf{KPI}} & \multicolumn{1}{c|}{\textbf{Description}} \\
\hline
\multicolumn{2}{|c|}{\textbf{Common Features}} \\
\hline
\textit{Cell Index} & Serving cell identifier. \\
\textit{Location} (3 Features) & Longitude, latitude, and altitude. \\
\textit{Velocity} $\&$ \textit{$\Delta$Velocity} & Current velocity of the UE (km/h). \\
\textit{RSRP} $\&$ \textit{$\Delta$RSRP} & Reference Signal Received Power (dBm). \\
\textit{RSRQ} $\&$ \textit{$\Delta$RSRQ} & Reference Signal Received Quality (dB). \\
\textit{RSSI} $\&$ \textit{$\Delta$RSSI} & Received Signal Strength Indicator (dBm). \\
\textit{SNR} $\&$ \textit{$\Delta$SNR} & Signal-to-Noise Ratio (dB). \\
\textit{CQI} $\&$ \textit{$\Delta$CQI} & Channel Quality Indicator. \\
\hline
\textit{HandoverFlag} & Handover Occurrence (1/0). \\
\hline
\multicolumn{2}{|c|}{\textbf{Beyond Throughput Features}} \\
\hline
\textit{TimeDifference} & Time elapsed for consecutive samples (s). \\
\textit{Distance} & Distance moved between samples (m). \\
\textit{NRxRSRP} $\&$ \textit{$\Delta$NRxRSRP} & Neighboring cell RSRP (dBm). \\
\textit{NRxRSRQ} $\&$ \textit{$\Delta$NRxRSRQ} & Neighboring cell RSRQ (dB). \\
\textit{BitrateDL, BitrateUL} &  Downlink and uplink throughput (kbps). \\
\textit{State} & Download state (I/D). \\
\hline
\multicolumn{2}{|c|}{\textbf{DoNext Features}} \\
\hline
\textit{TrackingAreaCode} & Tracking Area Code for base station. \\
\textit{TimingAdvance} & Timing Advance for the UE (s). \\
\textit{Bearing} & Current bearing angle for the UE ($^\circ$). \\
\textit{Location Accuracy}  & Measurement accuracy by the device. \\
\textit{Bearing Accuracy} & Measurement accuracy by the device. \\
\textit{Velocity Accuracy} & Measurement accuracy by the device. \\
\textit{SS-RSRP} $\&$ \textit{$\Delta$SS-RSRP} & RSRP of the Synchronization Signal. \\
\textit{SS-RSRQ} $\&$ \textit{$\Delta$SS-RSRQ} & RSRQ of the Synchronization Signal. \\
\textit{SS-SINR} $\&$ \textit{$\Delta$SS-SINR} & SINR of the Synchronization Signal. \\
\hline
\end{tabular}
\vspace{-15 pt}
\end{table}

\begin{table*}[t]
\centering
\caption{Performance Comparison of Models on Handover Datasets}
\vspace{-5 pt}
\label{tab:allResults}
\begin{threeparttable}
\resizebox{\textwidth}{!}{%
\begin{tabular}{|l|ccccccc|ccccccc|}
\hline
\multirow{2}{*}{\textbf{Method}} 
& \multicolumn{7}{c|}{\textbf{Beyond Throughput Dataset}} 
& \multicolumn{7}{c|}{\textbf{DoNext Dataset}}  \\
\cline{2-15}
& \multicolumn{2}{c}{\textbf{Acc. (Train/Test)}} & \textbf{AUC} & \textbf{Prec.} & \textbf{Rec.} & \textbf{F1-Score} & \textbf{Runtime [s]} 
& \multicolumn{2}{c}{\textbf{Acc. (Train/Test)}} & \textbf{AUC} & \textbf{Prec.} & \textbf{Rec.} & \textbf{F1-Score} & \textbf{Runtime [s]} \\
\hline
GAM         & 0.894 & 0.837 & 0.920 & 0.796 & 0.742 & 0.767 & 110.9
            & 0.813 & 0.779 & \textbf{0.867} & 0.733  & 0.679 & 0.705 & 222.5 \\
EBM         & \textbf{0.946} & 0.835 & 0.913 & 0.776  & 0.738 & 0.755 & \textbf{5.9}
            & \textbf{0.866} & 0.781 & 0.861 & 0.757  & 0.673 & 0.712 & \textbf{6.3} \\
GAMI-Net    & 0.846 & 0.826 & 0.908 & 0.785 & 0.715 & 0.747 & 327.6
            & 0.788 & 0.779 & 0.865 & 0.724  & \textbf{0.698} & \textbf{0.728} & 1660.2 \\
LSTM+SHAP   & 0.944 & \textbf{0.914} & \textbf{0.938} & \textbf{0.879}  & \textbf{0.862} & \textbf{0.870} & 13.8/1915.3
            & 0.842 & \textbf{0.804} & 0.865 & \textbf{0.792}  & 0.674 & 0.727 & 58.2/1932.7 \\
\hline
\end{tabular}
}
\begin{tablenotes}
\item *\textbf{Highlighted} metrics indicate the best ones.
\end{tablenotes}
\end{threeparttable}
\vspace{-15 pt}
\end{table*}

\subsubsection{Beyond Throughput Dataset} 
The Beyond Throughput (BT) dataset was collected from a major Irish operator \cite{commDataset}. Unlike datasets with basic performance indicators, the authors gathered a comprehensive set of KPIs to reflect realistic network conditions. The collection procedure covered multiple mobility scenarios and diverse application use cases to ensure robustness and representativeness. In this study, we use only samples from the driving scenario to observe HO events. Table \ref{tab:datasetProperties} summarizes the BT dataset. Given the limited sample size and relatively infrequent HOs, the subset comprises 2508 samples with a $34.81\%$ HO rate (i.e., 873 events). In total, 25 features are included, as detailed in Table \ref{tab:kpi}. A notable property of the BT dataset is that it reports performance metrics for the SC/NC, making it well-suited for HO analysis.

\subsubsection{DoNext Dataset}
The DoNext dataset was collected across Dortmund using fleet and public-order vehicles on geofenced routes for broad spatial coverage, complemented by a dedicated platform for long-duration, rail, and stationary measurements \cite{doNextDataset}. Together, these campaigns produced an open dataset including mobile, static, and rail scenarios. Beyond basic indicators, DoNext records key radio link quality KPIs and includes active measurements of data rate and latency. In this study, we restrict ourselves to the mobility use case with the associated data to capture HOs. Details are provided in Table \ref{tab:datasetProperties}. The dataset includes 59,472 HO events across 152,775 samples, yielding a HO ratio of $38.93\%$. The features are listed in Table \ref{tab:kpi}. A key limitation is that only SC metrics are available, which could constrain the handover analysis where simultaneous SC/NC information is required.

\section{Comparative Performance Analysis}
\label{sec:results}
This section presents the experimental procedure and evaluates the classification and interpretability performance \footnote{\url{https://github.com/afs-code/XAI-in-WirelessCommunications-HAEOTF}}.

\subsection{Experiment Design}
With the objective of on-the-fly interpretability, we formulate HO detection as a binary classification task to identify and explain the KPI conditions associated with HO decisions, rather than triggering or predicting future HOs. To solve this task, standard implementations of GAM, EBM, and GAMI-Net were adopted using the configurations reported in \cite{afs2}. An LSTM with 50 hidden units was also trained using Adam, binary cross-entropy, and early stopping, and its predictions were assessed using SHAP to obtain post-hoc feature contributions and global importance scores \cite{shap}. A consistent experimental protocol was applied across all models, where the fANOVA models were trained using 10,000 DoNext samples and 2,000 BT samples,selected according to available dataset sizes. LSTM inputs were generated with a sliding window of length $n_{\text{steps}}=10$. An 80/20 train-test split was used, with 20$\%$ of the training set reserved for validation. SHAP was evaluated on a subset of 100 samples. Random sampling was employed without stratification, while consistent sampling and data-splitting procedures were applied across all models. The train, validation, and test sets were kept exclusive to prevent data leakage, and all experiments were repeated over 10 independent random seeds to mitigate sampling variability and obtain robust average performance estimates. Lastly, all experiments were conducted using an NVIDIA RTX 4080 with 16 GB vRAM, Intel Core i9-13980HX, and 32 GB RAM. 

\subsection{Handover Detection Analysis}
For HO detection, the evaluation metrics recommended in \cite{ebmComparison} are adopted to ensure a consistent comparison across models. Table \ref{tab:allResults} summarizes the mean predictive and computational performance of the considered models over 10 independent runs. For the BT dataset, EBM achieves the highest training accuracy of $0.946$, whereas its test accuracy decreases to $0.835$, suggesting a degree of overfitting. In contrast, LSTM provides the strongest overall classification performance, attaining the highest test accuracy $(0.914)$, AUC $(0.938)$, precision $(0.879)$, recall $(0.862)$, and F1-score $(0.870)$. This improvement can be attributed to its ability to capture the temporal dependencies inherent in the HO process. Regarding computational efficiency, EBM records the lowest runtime of $5.9$ s, compared with $110.9$ s for GAM and $327.6$ s for GAMI-Net. EBM's efficient boosting procedure based on shallow decision trees enables substantially faster model fitting and explanation generation than the remaining models \cite{ebm}. This result is also consistent with \cite{ebmComparison}, where EBM was reported to provide an order-of-magnitude speedup over more complex neural architectures such as GAMI-Net. Although LSTM training requires only $13.8$ s, the subsequent SHAP analysis incurs an explanation time of $1915.3$ s, demonstrating the considerable computational post-hoc overhead.

Similar trends are observed for the DoNext dataset. EBM again achieves the highest training accuracy of $0.866$, although its test accuracy decreases to $0.781$. Among the evaluated models, LSTM attains the highest test accuracy $(0.804)$ and precision $(0.792)$, while GAM provides the highest AUC $(0.867)$. In contrast, GAMI-Net achieves the highest recall $(0.698)$ and F1-score $(0.728)$. Overall, the relatively small differences among the reported metrics indicate more balanced predictive performance across the four models. Regarding runtime, EBM remains the most efficient model at $6.3$ s, followed by GAM at $222.5$ s and GAMI-Net at $1660.2$ s. The LSTM requires $58.2$ s for training, whereas the subsequent explanation process of SHAP requires an additional $1932.7$ s. 

\begin{figure}
    \centering
    \includegraphics[width=\linewidth]{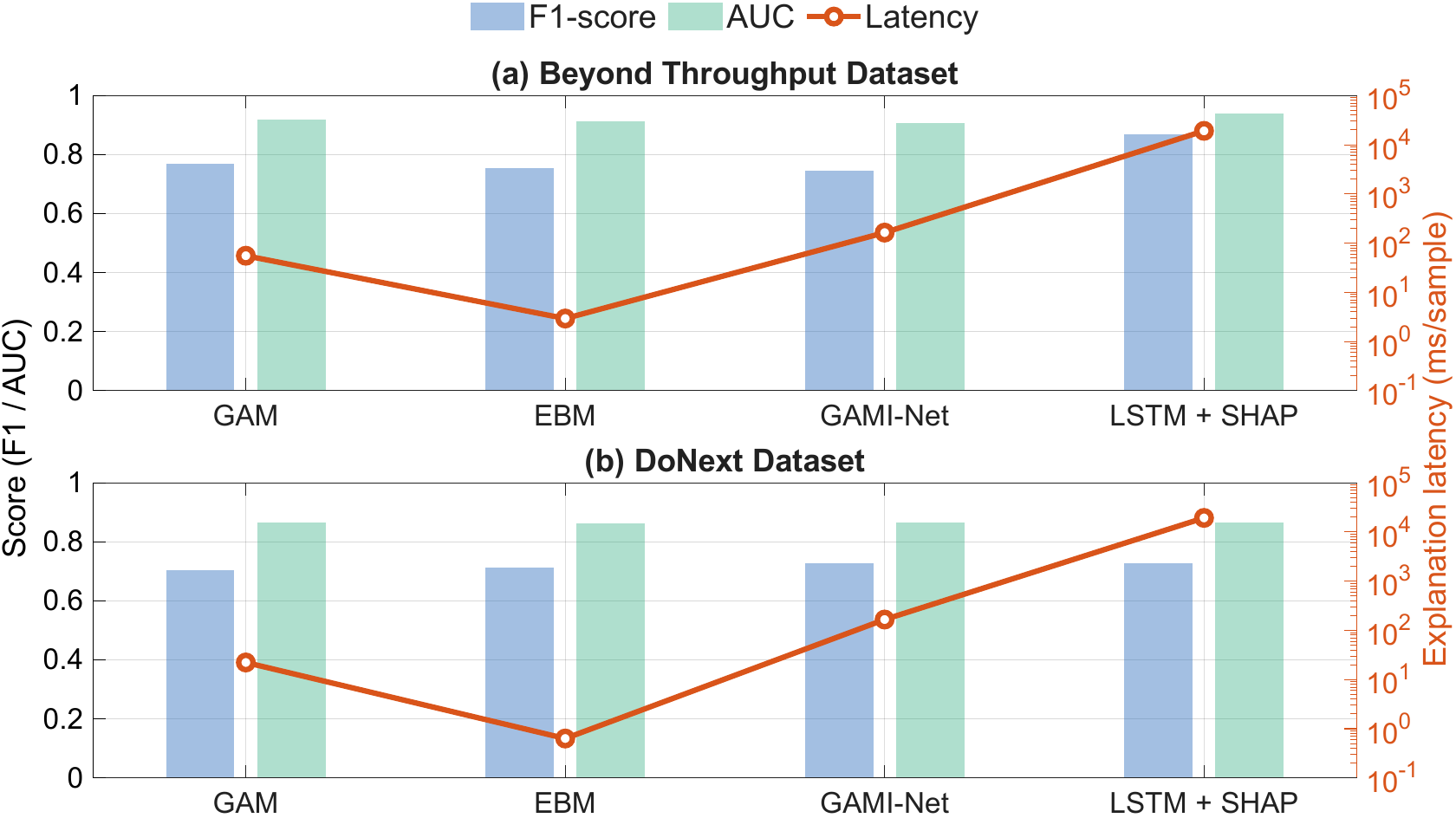}
    \caption{Accuracy-latency chart.}
    \label{fig:paretoChart}
    \vspace{-15 pt}
\end{figure}

The accuracy-latency trade-off is illustrated in Fig. \ref{fig:paretoChart}, where bar plots denote AUC and F1-score performances, and line plots indicate explanation latency measured in milliseconds per sample. The fANOVA-based models provide a substantial efficiency advantage for on-the-fly explanations, with EBM consistently achieving the lowest explanation latency. However, this efficiency comes at the cost of reduced predictive performance, particularly for the BT dataset. In contrast, LSTM+SHAP incurs significantly higher latency due to the post-hoc stage, despite its strong HO detection performance. These results highlight the operational trade-off between prediction and interpretability in vehicular networks.

We further examine the impact of dataset characteristics on model performance. BT achieves higher average AUC and F1-scores (\textbf{0.9198} and \textbf{0.7847}) than DoNext (\textbf{0.8644} and \textbf{0.7180}), respectively, suggesting the benefit of incorporating NC measurements for reliable HO detection.

\subsection{Explainability Analysis}
In addition to HO detection performance, we analyze explainability using feature-based ranking and relationship-based visualization. Feature importance is used to rank KPIs based on their contribution to HO decisions. The full set of KPIs is listed in Table \ref{tab:kpi}, while the most influential features are reported in Table \ref{tab:influentialFeatures}. HO decisions are primarily driven by channel-related metrics, particularly signal strength and quality indicators such as \textit{RSRP}, \textit{RSSI}, and \textit{RSRQ} \cite{handoverStandard, handoverMetricSelection}. While mobility-related features are also considered, broader aspects such as load management are beyond the scope of this study.

Table \ref{tab:influentialFeatures} presents the top-ranked KPIs identified by each model for the two datasets. Across all results, signal strength-related KPIs (e.g., \textit{RSRP}, \textit{RSRQ}, \textit{CQI}, and their variations) consistently dominate the HO decision process. EBM and GAMI-Net primarily select radio-related KPIs, whereas GAM and LSTM+SHAP occasionally include contextual features (e.g., \textit{Cell Index}, \textit{TrackingAreaCode}). This suggests that EBM and GAMI-Net more consistently capture physically meaningful indicators aligned with HO mechanisms. While topology-related KPIs (e.g., distance, identifiers) appear in some cases, their selection is less consistent \cite{handoverMetricSelection}. Notably, EBM uniquely highlights neighbor cell measurements (e.g., \textit{$\Delta$NRxRSRQ}), indicating its ability to capture inter-cell dynamics. Overall, inherently interpretable models provide clearer identification of key radio KPIs than post-hoc approaches.

\begin{table}[t]
\centering
\caption{Influential KPIs}
\vspace{-5 pt}
\label{tab:influentialFeatures}
\resizebox{0.5\textwidth}{!}{%
\begin{tabular}{|c|l|l|}
\hline
& {\textbf{Beyond Throughput}} & {\textbf{DoNext}}\\
\hline
\multirow{5}{*}{\textbf{GAM}} & \textit{Cell Index} & \textit{TrackingAreaCode} \\
& \textit{State} & \textit{SS-RSRP} \\
& \textit{Longitude} & \textit{Bearing} \\
& \textit{RSRP} & \textit{SS-RSRQ} \\
& \textit{BitrateUL} & \textit{RSRQ} \\
\hline
\multirow{5}{*}{\textbf{EBM}} & \textit{$\Delta$NRxRSRQ} & \textit{$\Delta$RSRP} \\
& \textit{$\Delta$RSRP} & \textit{$\Delta$CQI} \\
& \textit{Distance} & \textit{$\Delta$RSRQ} \\
& \textit{$\Delta$NRxRSRP} & \textit{$\Delta$Velocity} \\
& \textit{$\Delta$RSRP} x \textit{$\Delta$NRxRSRP} & \textit{TAC x $\Delta$RSRP} \\
\hline
\multirow{5}{*}{\textbf{GAMI-Net}} & \textit{Distance} & \textit{$\Delta$RSRP} \\
& \textit{RSRQ} & \textit{$\Delta$SINR} \\
& \textit{TimeDifference} & \textit{$\Delta$CQI} \\
& \textit{$\Delta$RSRQ} & \textit{$\Delta$RSRQ} \\
& \textit{$\Delta$RSRP} & \textit{$\Delta$Velocity} \\
\hline
\multirow{5}{*}{\textbf{LSTM + SHAP}} & \textit{Cell Index} & \textit{RSRQ} \\
& \textit{RSRQ} & \textit{CQI} \\
& \textit{CQI} & \textit{RSSI} \\
& \textit{BitrateDL} & \textit{$\Delta$RSRQ} \\
& \textit{$\Delta$CQI} & \textit{SINR} \\
\hline
\end{tabular}
}
\vspace{-15 pt}
\end{table}

\begin{figure*}[t]
    \centering
    \begin{subfigure}{0.49\columnwidth}
        \centering
        \includegraphics[width=\columnwidth]{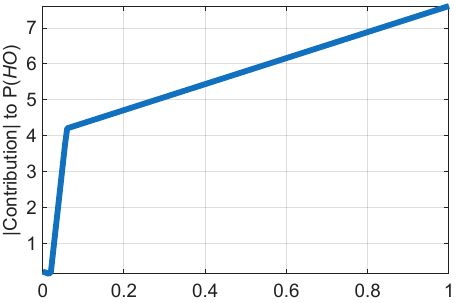}
        \caption{\textit{Distance}}
        \label{subfig:distance}
        \vspace{-3 pt}
    \end{subfigure}
    \hfill
    \begin{subfigure}{0.49\columnwidth}
        \centering
        \includegraphics[width=\columnwidth]{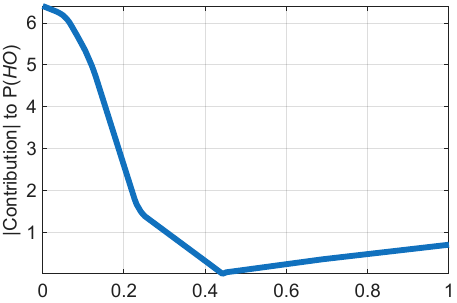}
        \caption{\textit{RSRQ}}
        \label{subfig:rsrq}
        \vspace{-3 pt}
    \end{subfigure}
    \hfill
    \begin{subfigure}{0.49\columnwidth}
        \centering
        \includegraphics[width=\columnwidth]{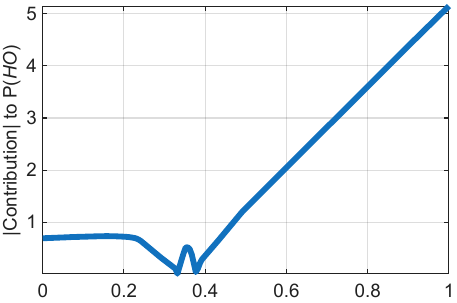}
        \caption{\textit{$\Delta$RSRQ}}
        \label{subfig:deltaRSRQ}
        \vspace{-3 pt}
    \end{subfigure}
    \hfill
    \begin{subfigure}{0.49\columnwidth}
        \centering
        \includegraphics[width=\columnwidth]{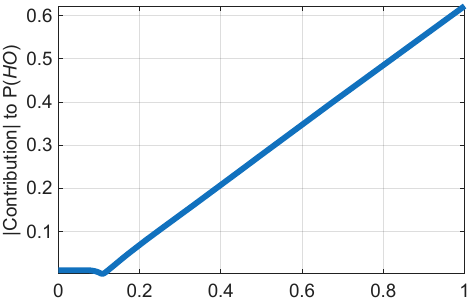}
        \caption{\textit{NRxRSRQ}}
        \label{subfig:nrRSRQ}
        \vspace{-3 pt}
    \end{subfigure}
    \caption{GAMI-Net visualizations for Distance and \textit{RSRQ}-related KPIs for the BT dataset.}
    \label{fig:rsrqVis}
    \vspace{-15 pt}
\end{figure*}

After identifying the most influential KPIs, we next focus on the visualization capability of inherently explainable models, which is largely absent in widely used post-hoc interpretability methods. In this context, we investigate the \textit{Distance} and \textit{RSRQ}-related KPIs present in the BT dataset from the perspective of both visualization and their underlying mathematical relationships. To motivate this analysis, we first recall the definition of the \textit{RSRQ} as specified by 3GPP as\cite{afs2}
\begin{equation}
    \textit{RSRQ} \;=\; \frac{N \times \textit{RSRP}}{\textit{RSSI}},
    \label{eq:RSRQ}
\end{equation}
where $N$ denotes the number of resource blocks within the measurement bandwidth used for \textit{RSSI} calculation. The \textit{RSRP} in a 5G network can be expressed as \cite{afs1}
\begin{equation}
    \textit{RSRP} = \frac{1}{N_{\mathrm{RE}}} \sum_{n=1}^{N_{\mathrm{RE}}} P_{\mathrm{i}}(n),
    \label{eq:rsrp}
\end{equation}
where $N_{\mathrm{RE}}$ is the number of resource elements (REs) carrying the reference signal (RS), and $P_{\mathrm{i}}(n)$ denotes the linear received power of the $n$-th RS. The received power per resource element can be written as
\begin{equation}
    P_{\mathrm{i}} = \frac{1}{K} \sum_{k=1}^{K} \left| r_{\mathrm{i}}[k] \right|^2,
    \label{eq:receivedSignalv1}
\end{equation}
where $K$ is the number of samples and $r_{\mathrm{i}}[k]$ is the received signal at the $k$-th sample. For analytical clarity, small-scale fading effects are omitted in this formulation. Furthermore, the received signal power can be related to the transmitter–receiver distance via the Friis transmission equation \cite{friisFormula}
\begin{equation}
    \left|r_{\mathrm{i}}\right|^2 \;\propto\; \frac{P_t}{(4 \pi d)^2},
    \label{eq:receivedSignalv2}
\end{equation}
where $P_t$ is the transmit power of the BS and $d$ denotes the distance between the UE and the BS. Here, frequency-dependent terms and antenna gains are omitted for simplicity. Hence, we can obtain the direct analytical association between \textit{Distance} and \textit{RSRQ} metrics from Equations \eqref{eq:RSRQ}–\eqref{eq:receivedSignalv2}. Lastly, we take inter-cell dynamics into account, which leads to a cause–and–effect interpretation of the HO event as follows
\begin{equation}
    d \uparrow \;\Rightarrow\; \textit{RSRQ}_{\mathrm{SC}} \downarrow 
    \;\land\; \textit{RSRQ}_{\mathrm{NC}} \uparrow 
    \;\Rightarrow\; \text{P}(HO) \uparrow,
    \label{eq:relationship}
\end{equation}
where $\textit{RSRQ}_{\mathrm{SC}}$ and $\textit{RSRQ}_{\mathrm{NC}}$ denote the RSRQ values of the SC and the NC, respectively, and $\text{P}(HO)$ defines the HO occurrence probability. The relationship in \eqref{eq:relationship} can be also interpreted geometrically by considering the UE on a two-dimensional plane between the SC and a NC, as shown in Fig. \ref{fig:SystemModel}. As the UE moves away from the SC, the propagation distance increases, leading to a degradation in $\textit{RSRQ}_{\mathrm{SC}}$, while the proximity to the NC improves $\textit{RSRQ}_{\mathrm{NC}}$. This opposing behavior directly supports HO triggering conditions such as the A3 event in \eqref{eq:A3}, where a HO is initiated once the NC quality exceeds that of the SC. Thus, it increases the likelihood of HO occurrence as the UE continues its movement.

To further validate and visualize this cause–effect relationship, Fig. \ref{fig:rsrqVis} presents the learned contributions of key KPIs using GAMI-Net \cite{gaminet}. The results show that \textit{Distance} exhibits a monotonic increasing relationship with $\text{P}(HO)$, consistent with signal attenuation models where received power decays with squared distance. Among radio metrics, \textit{RSRQ} demonstrates a negative relationship with HO probability, indicating that lower signal quality increases the likelihood of HO. In contrast, \textit{$\Delta$RSRQ} captures sensitivity to temporal variations, where larger changes contribute more to HO decisions. Similarly, \textit{NRxRSRQ} shows a positive relationship with $\text{P}(HO)$, confirming that stronger NC quality promotes HO triggering. Together, these results reveal how different system parameter configurations influence the HO process, while remaining consistent with the analytical relationship in \eqref{eq:relationship}.

\section{Conclusion}
\label{conclusion}
This paper investigated HO detection in vehicular 5G NR networks through an explainability-on-the-fly lens by comparing inherently interpretable fANOVA models with an LSTM baseline using post-hoc SHAP explanations. While the LSTM achieved strong detection performance, its explainability incurred significant computational overhead, limiting its suitability for latency-critical HO scenarios. In contrast, fANOVA-based models provided competitive detection performance with orders-of-magnitude lower explanation latency. Feature ranking analysis confirmed that intrinsically explainable models consistently identify physically meaningful radio link KPIs, such as \textit{RSRP}, \textit{RSRQ}, and their temporal variations. Additionally, visualization analysis revealed interpretable cause-and-effect relationships that align with the mathematically derived HO decision relationship. Overall, the results demonstrate that fANOVA-based models enable efficient and transparent HO detection in next-generation mobile networks.

\section*{Acknowledgment}
The authors acknowledge using ChatGPT to refine the grammar and enhance the English language expressions.

\bibliographystyle{IEEEtran}
\bibliography{refer}

\end{document}